\documentclass[a4paper, 10pt, journal]{ieeeconf}  

\usepackage{amsmath}
\usepackage{amsfonts}
\usepackage{float}
\usepackage{algorithm}
\usepackage{algpseudocode}
\usepackage{booktabs}
\usepackage{siunitx}  
\usepackage{multirow}
\usepackage{dashrule}
\usepackage{graphicx}

\title{\LARGE \bf Controllability-Informed Distributed Electric Propulsion Aircraft Design Using Probabilistic Attainable Moment Sets}

\author{Umut Demir$^{1}$, Torbj\o rn Cunis$^{1}$, Aamir Ahmad$^{1}$ and Walter Fichter$^{1}$
\thanks{$^{1}$University of Stuttgart, Faculty of Aerospace Engineering and Geodesy, Institute of Flight Mechanics and Control (iFR), Germany.
        {\tt\small firstname.lastname@ifr.uni-stuttgart.de}}}

\begin{document}

\maketitle
\thispagestyle{empty}
\pagestyle{empty}

\begin{abstract}
Integrated aircraft design combines models of aerodynamics, propulsion, mass, and mission performance in one design loop, so that the disciplines can be traded against each other from the earliest stage. For configurations with distributed electric propulsion (DEP), part of the control authority comes from the propulsion system itself, which directly affects the required size of the tail and should therefore be known while the aircraft is being sized. This paper presents a controllability measure that is evaluated for every design in the loop and returned to it as a design signal. The attainable moment set of each candidate design is computed from the surrugate models of the loop and it is compared with a required moment set. Since the low-fidelity surrugate models are uncertain, the comparison is formulated with probabilistic attainable moment sets: the uncertainty is propagated to the boundary of the set, which yields a margin for every direction in the moment space and the probability of violating the requirement. The design loop thereby informed whether a configuration provides sufficient control authority, on which axes it falls short, and how this changes with the design variables. The approach is demonstrated on a DEP regional aircraft, where the signal is used to assess the size of the vertical tail.
\end{abstract}

\section{INTRODUCTION}\label{sec:introduction}
Distributed electric propulsion (DEP) replaces a few large propulsion units by many smaller ones that are distributed over the airframe. The propulsion system can be integrated closely with the wing, which increases the maximum lift coefficient, permits a smaller wing that is better matched to cruise, and adds redundancy~\cite{kim2018review}.. The same propulsion units can also generate moments: differential thrust acts as an additional control effector, in particular about the yaw axis, where it has been studied as a means of yaw control and of reducing the vertical tail~\cite{vechtel2022aspects,nguyen2018reduction}. These effects do not belong to a single discipline. A smaller wing changes mass and trim, the propulsion layout changes the available control authority, and a smaller tail changes drag, mass, and balance, so that the configuration has to be evaluated as a whole. This is the purpose of integrated aircraft design loops, in which models of aerodynamics, propulsion, mass, and mission performance are iterated until the aircraft is consistent, and which have been extended to hybrid-electric and fuel-cell electric configurations with distributed propulsion~\cite{mangold2023preliminary,schlittenhardt2025preliminary}. In the collaborative research center SynTrac, such a loop is the common platform to which the individual disciplines contribute models and surrogates in order to exploit the synergies between them. Controllability is not yet represented in it: the loop has no information on whether a design provides sufficient control authority, although for a DEP aircraft this authority depends on the very quantities that the loop is sizing.

A DEP aircraft is overactuated: its propulsion units and aerodynamic control surfaces together provide more effectors than the three rotational axes to be controlled. Control allocation decides how a commanded moment is distributed among these effectors, but it can only deliver moments that the effectors are able to produce within their limits. The set of these moments is fixed by the sizing and placement of the effectors, that is, by aircraft design, and it bounds everything that allocation and the control laws can achieve in flight. To assess controllability, the attainable moment set (AMS) was introduced by Durham in the context of constrained control allocation as the set of moments that a redundant suite of effectors can generate within its position limits~\cite{durham1993constrained,durham1994three,durham1994attainable}. For a linear effectiveness model it is a convex polytope whose geometry and computation are well understood~\cite{durham2017aircraft,johansen2013control}. Beyond allocation, the AMS has become a tool to visualize and quantify control authority during vehicle design~\cite{pei2018generic}. Its size alone, however, does not state whether the available authority is sufficient. S\"opper et al.~\cite{sopper2021required} therefore introduced the required moment set (RMS), which is assembled from the aerodynamic disturbances over the flight envelope and the maneuver accelerations demanded by the mission, and assessed controllability through the margins between the two sets along a common grid of query directions. Zhang et al.~\cite{zhang2021attainable} turned these margins into a cost function and optimized the propeller installation angles of an eVTOL such that its AMS covers the RMS. Both works are deterministic, and the need to include model uncertainty in order to obtain probabilistic controllability statements was already noted in~\cite{sopper2021required}. Tsagkaris et al.~\cite{tsagkaris2026probabilistic} recently addressed it with the probabilistic attainable moment set (PAMS), which propagates parametric uncertainty of the effector model to the boundary of the set through a first-order expansion along fixed rays and uses the probability of violating the requirement as the objective for optimizing multirotor tilt angles.

In this work, the PAMS is adopted as the controllability measure and integrated into the aircraft sizing loop~\cite{schlittenhardt2025preliminary}. Every candidate configuration that the loop produces is analyzed with the aerodynamic tool AVL~\cite{drela2010avl}, which provides the control effectiveness of the converged design. The uncertainty of these low-fidelity predictions is propagated to the attainable set, and the resulting PAMS is compared with a required moment set. The comparison quantifies each candidate by its margin in every direction of the moment space and by the probability of violating the requirement, and this information is returned to the sizing loop. Because the measure is evaluated on the converged design, it reflects the changes in mass, center of gravity, inertia, and trim state that a change of the effectors causes. The approach is demonstrated on a DEP regional aircraft.

\section{Problem Formulation}\label{sec:problem_formulation}
This section introduces the three sets on which the controllability measure is built. The attainable moment set describes what the effectors of a system can generate, its probabilistic extension accounts for uncertainty in the underlying models, and the required moment set states what the system must be able to generate. The comparison of the attainable with the required set then yields the probability that the requirements are violated.

\subsection{Attainable Moment Set}\label{subsec:ams}

Following~\cite{tsagkaris2026probabilistic}, consider a system whose dynamics separate into a state-dependent and an input-dependent contribution,
\begin{equation}
    \dot{\mathbf{x}} = f(\mathbf{x}) + g(\mathbf{u}),
    \label{eq:dynamics}
\end{equation}
where $\mathbf{x} \in \mathbb{R}^{n_m}$ is the state vector, $\mathbf{u} \in \mathbb{R}^{n_u}$ the vector of control inputs, and $f: \mathbb{R}^{n_m} \rightarrow \mathbb{R}^{n_m}$ and $g: \mathbb{R}^{n_u} \rightarrow \mathbb{R}^{n_m}$ are continuous functions. This form is less restrictive than the linear effectiveness models commonly adopted in the AMS literature~\cite{durham1993constrained,durham2017aircraft}. The contribution of the effectors,
\begin{equation}
    \boldsymbol{\nu} = g(\mathbf{u}) \in \mathbb{R}^{n_m},
    \label{eq:gen_moment}
\end{equation}
is referred to as the generalized moment. Let the admissible control set
\begin{equation}
    \mathcal{U} \subseteq \mathbb{R}^{n_u}
    \label{eq:acs_general}
\end{equation}
contain all physically possible combinations of the control inputs. The attainable moment set is the image of the admissible control set under $g$,
\begin{equation}
    \mathcal{A} = \left\{ \boldsymbol{\nu} \in \mathbb{R}^{n_m} \;\middle|\; \boldsymbol{\nu} = g(\mathbf{u}),\; \mathbf{u} \in \mathcal{U} \right\} = g(\mathcal{U}).
    \label{eq:ams_general}
\end{equation}
The state is commonly defined in terms of linear and angular velocities, $\mathcal{A}$ is a set of accelerations.

\subsection{Probabilistic Attainable Moment Set}\label{subsec:pams}

The definition~\eqref{eq:ams_general} assumes that the input map $g$ and the admissible control set $\mathcal{U}$ are known exactly. Following~\cite{tsagkaris2026probabilistic}, let the system instead depend on a vector of $n_p$ uncertain parameters, modeled as an uncorrelated normal random vector,
\begin{equation}
    \mathbf{p} \sim \mathcal{N}(\boldsymbol{\mu}_p, \boldsymbol{\Sigma}_p), \qquad \boldsymbol{\Sigma}_p = \mathrm{diag}(\sigma_1^2, \ldots, \sigma_{n_p}^2).
    \label{eq:param_dist}
\end{equation}
The dynamics~\eqref{eq:dynamics} are extended to
\begin{equation}
    \dot{\mathbf{x}} = f(\mathbf{x}) + g(\mathbf{u}, \mathbf{p}),
    \label{eq:dynamics_p}
\end{equation}
so that the generalized moment $\boldsymbol{\nu} = g(\mathbf{u}, \mathbf{p})$ is a random vector, while $f(\mathbf{x})$ remains deterministic. The parameters may also act on the effector limits, $\mathcal{U} = \mathcal{U}(\mathbf{p})$. The attainable moment set then becomes a random set,
\begin{equation}
    \mathcal{A}(\mathbf{p}) = \left\{ \boldsymbol{\nu} \in \mathbb{R}^{n_m} \;\middle|\; \boldsymbol{\nu} = g(\mathbf{u}, \mathbf{p}),\; \mathbf{u} \in \mathcal{U}(\mathbf{p}) \right\},
    \label{eq:ams_p}
\end{equation}
of which the deterministic set~\eqref{eq:ams_general} is the realization at $\mathbf{p} = \boldsymbol{\mu}_p$. For a prescribed confidence level $\alpha \in (0, 1)$, the $\alpha$-confidence probabilistic attainable moment set (PAMS) is defined as
\begin{equation}
    \mathcal{A}^{(\alpha)} = \left\{ \boldsymbol{\nu} \in \mathbb{R}^{n_m} \;\middle|\; \mathbb{P}\big( \boldsymbol{\nu} \in \mathcal{A}(\mathbf{p}) \big) \geq \alpha \right\},
    \label{eq:pams}
\end{equation}
that is, the set of generalized moments that the effectors can generate with a probability of at least $\alpha$. The sets are nested, $\mathcal{A}^{(\alpha_2)} \subseteq \mathcal{A}^{(\alpha_1)}$ for $\alpha_1 \leq \alpha_2$, so a higher confidence level yields a smaller set.

\subsection{Required Moment Set}\label{subsec:rms}

The AMS and the PAMS describe what the effectors can generate, but not whether this is sufficient. Following~\cite{sopper2021required}, sufficiency is expressed by a required moment set (RMS) $\mathcal{R} \subset \mathbb{R}^{n_m}$, which contains all generalized moments that the system must be able to generate in order to fulfill its requirements. In~\cite{sopper2021required} this set is assembled from disturbance and maneuver requirements over the flight envelope; where such requirements are not available, it can instead be taken from a reference system whose control authority is known to be adequate, which is the approach followed in this work. In both cases $\mathcal{R}$ is deterministic, and the requirements are fulfilled if the required set is contained in the attainable set,
\begin{equation}
    \mathcal{R} \subseteq \mathcal{A}.
    \label{eq:containment}
\end{equation}

For the random set $\mathcal{A}(\mathbf{p})$ in~\eqref{eq:ams_p}, the containment~\eqref{eq:containment} is a random event that holds for some realizations of the uncertain parameters and fails for others. The quantity of interest is therefore the probability that the requirements are violated~\cite{tsagkaris2026probabilistic},
\begin{equation}
    \varepsilon = 1 - \mathbb{P}\big( \mathcal{R} \subseteq \mathcal{A}(\mathbf{p}) \big).
    \label{eq:eps_set}
\end{equation}
A value of $\varepsilon$ close to zero means that the requirements are met for almost all realizations of the parameters, and a value close to one that they are almost certainly violated. Unlike the deterministic test~\eqref{eq:containment}, $\varepsilon$ accounts for how far the boundary of the attainable set lies from that of the required set relative to its uncertainty.

\section{Evaluation of the Controllability Measure}\label{sec:evaluation}

This section describes how the controllability measure is obtained for one  design iteration of the sizing loop. The aircraft is first trimmed, its attainable set is constructed from the trimmed model, the uncertainty of the model is propagated to the boundary of this set, and the result is compared with the required set to obtain the violation probability~\eqref{eq:eps_set}. The reference aircraft, which defines the required set, is treated with the same steps.

\subsection{Aerodynamic Analysis with AVL}
The aerodynamic quantities of the method are computed with the Athena Vortex Lattice (AVL) program~\cite{drela2010avl}, an extended vortex-lattice method for the aerodynamic and flight-dynamic analysis of rigid aircraft of arbitrary configuration. It models the lifting surfaces with horseshoe vortices and returns forces, moments, and their stability and control derivatives from one linear, inviscid solution. Its low cost and its direct output of control derivatives make it a natural choice for conceptual design, where the geometry changes with every design candidate.

\begin{figure}[t]
    \centering
    \includegraphics[width=\columnwidth]{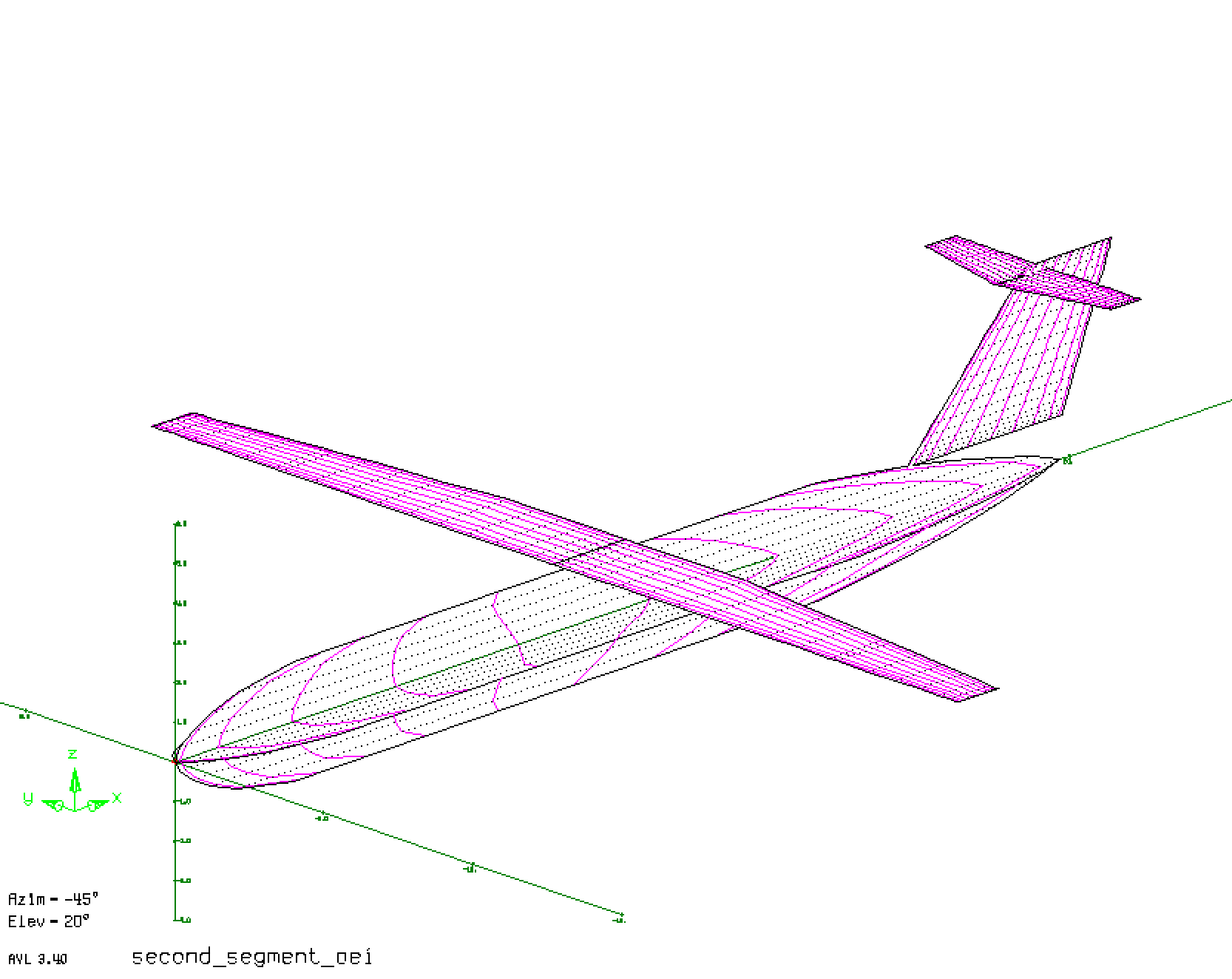}
    \caption{AVL model of the reference aircraft. The wing and the T-tail are modeled as lifting surfaces, and the fuselage is represented by a horizontal and a vertical flat surface. Axes in meters.}
    \label{fig:avl_reference}
\end{figure}

\subsection{Trim}

The  aircraft is trimmed in AVL at the Mach number and altitude of interest with its take-off mass $m$. AVL pairs each operating variable with one constraint~\cite{drela2010avl}: the angle of attack $\alpha_0$ is solved for the lift that balances the weight and the elevator deflection for zero pitching moment,
\begin{equation}
    C_L = \frac{m g}{\bar{q} S}, \qquad C_m = 0,
    \label{eq:trim}
\end{equation}
\subsection{Attainable Set}
The effectors are the control surfaces and the $n_T$ propellers. For each control surface, AVL returns the derivatives of the rolling, pitching, and yawing moment coefficients with respect to its deflection at the trimmed state. Multiplied by the dynamic pressure, the reference area, and the reference lengths, they give the moment per degree of deflection and form one column of the effectiveness matrix each,
\begin{equation}
    \mathbf{B}_\delta = \bar{q} S\, \mathrm{diag}(b, \bar{c}, b) \left. \frac{\partial \begin{bmatrix} C_l & C_m & C_n \end{bmatrix}^\top}{\partial \boldsymbol{\delta}} \right|_{\alpha_0,\, \boldsymbol{\delta}_0}.
    \label{eq:B_aero}
\end{equation}
The propellers are not part of the AVL model. Each one adds a column with its moment per unit thrust, which follows from its position $\mathbf{r}_i = [x_i \;\, y_i \;\, z_i]^\top$ relative to the center of gravity and the inclination $\iota_T$ of the thrust line,
\begin{equation}
    \mathbf{b}_{T,i} = \begin{bmatrix} -y_i \sin\iota_T \\ z_i \cos\iota_T + x_i \sin\iota_T \\ -y_i \cos\iota_T \end{bmatrix}.
    \label{eq:B_thrust}
\end{equation}
The columns are collected in $\mathbf{B} = [\mathbf{B}_\delta \;\, \mathbf{b}_{T,1} \cdots \mathbf{b}_{T,n_T}]$ and converted to angular accelerations with the inertia tensor $\mathbf{J}$ of the converged design,
\begin{equation}
    \mathbf{G} = \mathbf{J}^{-1} \mathbf{B}.
    \label{eq:accel_mapping}
\end{equation}

The admissible control set contains the effector increments that remain available about trim,
\begin{equation}
    \mathcal{U} = \left\{ \Delta\mathbf{u} \;\middle|\; \mathbf{u}_{\min} - \mathbf{u}_0 \leq \Delta\mathbf{u} \leq \mathbf{u}_{\max} - \mathbf{u}_0 \right\}.
    \label{eq:acs}
\end{equation}
For a control surface, these are its deflection limits minus the trim deflection. A propeller can reduce its thrust from the trim value $T_0$ to zero and increase it to the maximum thrust available at the trim airspeed $V$,
\begin{equation}
    T_{\max} = \min\left\{ \frac{\eta_p P}{V},\; \left( 2 \rho A_p\, (\eta_p P)^2 \right)^{1/3} \right\},
    \label{eq:tmax}
\end{equation}
with the power $P$, the propeller efficiency $\eta_p$, and the disk area $A_p$.

The attainable set is
\begin{equation}
    \mathcal{A} = \left\{ \boldsymbol{\nu} = \mathbf{G}\, \Delta\mathbf{u} \;\middle|\; \Delta\mathbf{u} \in \mathcal{U} \right\}.
    \label{eq:ams}
\end{equation}
Since $\mathcal{U}$ is the convex hull of its $2^{n_u}$ vertices and the mapping is linear, $\mathcal{A}$ is the convex hull of the mapped vertices~\cite{durham2017aircraft}.

Under uncertainty, this construction is repeated for realizations of the parameters. The parameters $\mathbf{p}$ represent the uncertainty of the low-fidelity models from which the set is built, and they are of two kinds. The aerodynamic parameters are factors on the control effectiveness predicted by AVL, one per control surface, and the inertia parameters are factors on the moments of inertia obtained from the mass build-up of the sizing loop. All factors have unit mean, so that the nominal set corresponds to the model predictions. A sample is drawn as
\begin{equation}
    \mathbf{p}^{(i)} = \boldsymbol{\mu}_p + \boldsymbol{\sigma} \circ \hat{\mathbf{p}}^{(i)}, \qquad \hat{\mathbf{p}}^{(i)} \sim \mathcal{N}(\mathbf{0}, \mathbf{I}), \qquad i = 1, \ldots, N,
    \label{eq:mc_sample}
\end{equation}
where $\boldsymbol{\sigma}$ collects the standard deviations of the parameters and $\circ$ denotes the element-wise product. The aerodynamic factors in $\mathbf{p}^{(i)}$ scale the corresponding columns of $\mathbf{B}_\delta$, and the inertia factors scale the inertia tensor $\mathbf{J}$, which together give $\mathbf{G}(\mathbf{p}^{(i)}) = \mathbf{J}(\mathbf{p}^{(i)})^{-1} \mathbf{B}(\mathbf{p}^{(i)})$. The vertices of $\mathcal{U}$ are mapped through this matrix, and their convex hull is one realization of the random set~\eqref{eq:ams_p},
\begin{equation}
    \mathcal{A}^{(i)} = \left\{ \boldsymbol{\nu} = \mathbf{G}(\mathbf{p}^{(i)})\, \Delta\mathbf{u} \;\middle|\; \Delta\mathbf{u} \in \mathcal{U} \right\}.
    \label{eq:ams_sample}
\end{equation}

Sets cannot be averaged or compared directly, so each set is described by its radii. A radius is the distance from the trim point, which is the origin of the moment space, to the boundary of the set along a given direction; it is the largest angular acceleration that the effectors can produce in that direction. The directions are unit vectors
\begin{equation}
    \hat{\mathbf{n}}(\boldsymbol{\phi}) = \begin{bmatrix} \cos\phi_1 & \sin\phi_1 \cos\phi_2 & \sin\phi_1 \sin\phi_2 \end{bmatrix}^\top,
    \label{eq:direction}
\end{equation}
taken from a grid $\Phi$ of the angles $\boldsymbol{\phi} = (\phi_1, \phi_2)$ that covers the sphere and is fixed once for all sets. To evaluate the radius, the convex hull is written in half-space form $\mathcal{A} = \{ \boldsymbol{\nu} \mid \mathbf{H} \boldsymbol{\nu} \leq \mathbf{h} \}$, where each row $\mathbf{H}_j$ is the outward unit normal of a facet and $h_j$ its distance from the origin. A point $r\, \hat{\mathbf{n}}$ on the ray lies in the set as long as it is on the inner side of every facet, and the radius is the distance at which the first facet is reached,
\begin{equation}
    r_{\mathcal{A}}(\boldsymbol{\phi}) = \min_{j:\, \mathbf{H}_j \hat{\mathbf{n}} > 0} \frac{h_j}{\mathbf{H}_j \hat{\mathbf{n}}(\boldsymbol{\phi})}.
    \label{eq:radius}
\end{equation}
Since the set is convex and contains the origin, every ray leaves it exactly once, so the radii over all directions describe the set completely.

Applied to the realizations~\eqref{eq:ams_sample}, this gives one radius $r^{(i)}(\boldsymbol{\phi})$ per sample and direction. Because the directions are the same for every sample, the uncertainty of the set appears as the scatter of a scalar in each direction, with mean and standard deviation
\begin{align}
    \bar{r}(\boldsymbol{\phi}) &= \frac{1}{N} \sum_{i=1}^{N} r^{(i)}(\boldsymbol{\phi}), \qquad \\
    \sigma_r^2(\boldsymbol{\phi}) &= \frac{1}{N-1} \sum_{i=1}^{N} \left( r^{(i)}(\boldsymbol{\phi}) - \bar{r}(\boldsymbol{\phi}) \right)^2.
    \label{eq:mc_stats}
\end{align}

\begin{figure}[t]
    \centering
    \includegraphics[width=0.9\columnwidth]{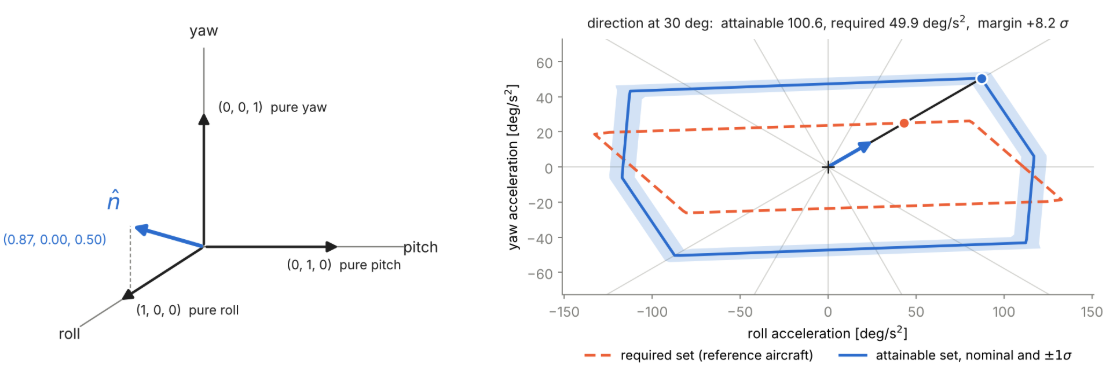}
    \caption{Direction and radius. Left: a direction $\hat{\mathbf{n}}$ is a unit vector in the space of roll, pitch, and yaw acceleration. Right: cut through the roll--yaw plane for the DEP aircraft with $c_{VT} = 0.119$; the radius along $\hat{\mathbf{n}}$ is the distance from the trim point to the boundary, here $101$~deg/s$^2$ for the attainable set and $50$~deg/s$^2$ for the required set.}
    \label{fig:direction_radius}
\end{figure}

The required set is described by its radii $r_{\mathcal{R}}(\boldsymbol{\phi})$ on the same grid, so that the two sets can be compared direction by direction. Fig.~\ref{fig:direction_radius} illustrates both quantities: a direction is one combination of roll, pitch, and yaw acceleration, and the radius is the distance from the trim point to the boundary of a set along that direction.

The comparison of the two sets is expressed in each direction as a margin, the distance between the mean attainable radius and the required radius in units of the standard deviation,
\begin{equation}
    z(\boldsymbol{\phi}) = \frac{\bar{r}(\boldsymbol{\phi}) - r_{\mathcal{R}}(\boldsymbol{\phi})}{\sigma_r(\boldsymbol{\phi})}, \qquad z_{\min} = \min_{\boldsymbol{\phi} \in \Phi} z(\boldsymbol{\phi}).
    \label{eq:sigma_margin}
\end{equation}
A positive margin means that the requirement is covered in that direction, and its value states by how many standard deviations. The direction that attains the minimum margin $z_{\min}$ is the binding direction; it identifies the axis, or combination of axes, on which the design is closest to violating the requirement. The violation probability~\eqref{eq:eps_set} considers all directions jointly. A sample violates the requirement if its radius falls below the required one in at least one direction, and the probability is estimated by the fraction of such samples,
\begin{equation}
    \varepsilon \approx \frac{1}{N} \sum_{i=1}^{N}  \left[ \min_{\boldsymbol{\phi} \in \Phi} \left( r^{(i)}(\boldsymbol{\phi}) - r_{\mathcal{R}}(\boldsymbol{\phi}) \right) < 0 \right].
    \label{eq:eps_mc}
\end{equation}
The margin $z_{\min}$ and the probability $\varepsilon$ are the two outputs of the controllability measure. The margin varies smoothly with the design and remains informative when $\varepsilon$ is close to zero or one, and it is therefore the quantity that is passed to the optimization.

\subsection{Bayesian Optimization of the Tail Size}\label{sec:bo}

The controllability measure becomes useful to the design only if it can steer a design variable, here the size of the vertical tail. Finding the best tail size is an optimization problem with three properties that rule out standard optimizers. First, every evaluation is expensive: a candidate has to be sized to convergence, analyzed aerodynamically, and trimmed before its mass and its control authority are known, which takes several minutes, so that only a few tens of candidates can be evaluated. Second, no gradients are available. Third, the constraints are as expensive as the objective, because they come from the same evaluation. Bayesian optimization is designed for this situation~\cite{jones1998efficient,gardner2014bayesian}. It builds a statistical model of the objective and of each constraint from the candidates evaluated so far, and uses these models to choose the next candidate where an improvement is most likely and the constraints are likely to hold. Each expensive evaluation is thereby placed where it is expected to be most informative.

\subsection{Problem}

The optimization takes the vertical tail volume coefficient as its design variable, $\theta = c_{VT}$, together with its bounds $\theta_{\min}$ and $\theta_{\max}$. For a given $\theta$, one evaluation of the sizing loop and the controllability analysis returns four quantities of the converged design: the take-off mass $m_{TO}$, the minimum margin $z_{\min}$ between the attainable and the required set, the directional stability $C_{n_\beta}$, and the trim angle of attack $\alpha_0$. The optimization seeks the lightest aircraft that covers the required set, remains directionally stable, and can be trimmed within the valid range of the aerodynamic model,
\begin{equation}
    \begin{aligned}
        \min_{\theta \in [\theta_{\min},\, \theta_{\max}]} \quad & m_{TO}(\theta) \\
        \text{subject to} \quad & z_{\min}(\theta) \geq 0, \\
        & C_{n_\beta}(\theta) \geq C_{n_\beta,\min}, \\
        & \alpha_0(\theta) \leq \alpha_{\max}.
    \end{aligned}
    \label{eq:bo_problem}
\end{equation}
Its output is the tail size $\theta^\star$ of the best design that satisfies all constraints, together with the mass of that design, its margins, and the constraint that limits it. If no evaluated design is feasible, the output is the design that comes closest to satisfying the constraints, which shows whether the design variable can influence the limiting constraint at all.

\subsection{Aircraft Sizing Loop}\label{subsec:sizing}

Both aircraft are designed with the extended version of SUAVE~\cite{lukaczyk2015suave} described by Schlittenhardt et al.~\cite{schlittenhardt2025preliminary}, which adds an iterative sizing loop and a fuel-cell electric energy network to the original environment. The mass and parasite-drag models follow handbook methods and are calibrated with component mass factors and segment-wise drag corrections against a reverse-engineered ATR~72-500, for which the calibrated model reproduces the take-off and operating empty mass within 0.5\,\%~\cite{schlittenhardt2025preliminary}. Wing loading and power loading are set by a sizing chart from the approach speed, take-off field length, cruise, and service ceiling requirements. The energy network represents the fuel cells, power converters, electric motors, propellers, thermal circuit, and liquid hydrogen storage as separate components; their masses follow from their rated power, and the tank is placed in the fuselage behind the cabin, so that the fuselage length depends on the hydrogen mass.

For a given set of top-level requirements, the loop iterates the take-off, operating empty, and fuel mass, the mission energy including reserves, the installed power, the hydrogen tank length, and the longitudinal wing position that places the center of gravity at its prescribed location, until these quantities are mutually consistent. The mission is solved in each iteration by enforcing force equilibrium at every mission point, with the hydrogen mass flow returned by the energy network. A change of one component therefore propagates to the others; a heavier or longer aircraft requires more power and fuel, which in turn increases the mass again~\cite{schlittenhardt2025preliminary}.

The tail surfaces are part of this iteration. Their areas are determined from the volume coefficients and the current lever arms. Since the wing position and the center of gravity move during the iteration, the same $c_{VT}$ leads to a different tail area for a different aircraft, and a change of $c_{VT}$ alters mass, balance, and inertia of the converged design. The converged design supplies the quantities that the attainable moment set requires: the geometry for the vortex-lattice model, the mass and the center of gravity for the trim, the component masses and positions for the inertia tensor, and the propeller positions and rated power for the thrust effectors.

\section{Results}\label{sec:results}
The method is applied to a pair of 70-seat regional aircraft of the ATR~72 class with hydrogen fuel-cell electric propulsion, both modeled and sized in SUAVE~\cite{lukaczyk2015suave}. The reference aircraft is a conventional twin with one propeller on each wing. The DEP aircraft replaces the two propulsion units by eight smaller ones that are distributed along the span, with the outermost propeller at the wing tip. Both aircraft share the top-level requirements in Table~\ref{tab:tlar}, the fuselage, the wing span, the T-tail arrangement, and the tail planform parameters, and both are sized to convergence with the same loop, so that mass, wing area, center of gravity, and inertia are consistent with the respective configuration. The results are presented in two parts. First, the probabilistic attainable moment set of the DEP aircraft is compared with the required moment set obtained from the reference aircraft, which shows on which axes the DEP aircraft covers the requirement and with what margin. Second, this comparison is used as a constraint in the Bayesian optimization of Section~\ref{sec:bo}, which determines the size of the vertical tail.

\begin{table}[t]
    \centering
    \caption{Top-level requirements of both aircraft}\label{tab:tlar}
    \begin{tabular}{ll}
        \toprule
        Passengers & 70 \\
        Design payload & 6720 kg \\
        Design range & 1169 km (631 nmi) \\
        Cruise speed & 136.3 m/s (265 kt TAS) \\
        Cruise altitude & 6706 m (22\,000 ft) \\
        \bottomrule
    \end{tabular}
\end{table}

Table~\ref{tab:aircraft} lists the converged data of both aircraft. The DEP design shown there is used for the comparison of the attainable and required sets in the first part; in the second part its vertical tail volume coefficient is varied by the optimization.

\begin{table}
    \centering
    \caption{Converged data of the reference and the DEP aircraft.}\label{tab:aircraft}
    \begin{tabular}{lcc}
        \toprule
        & Reference & DEP \\
        \midrule
        \multicolumn{3}{l}{\textit{Geometry}} \\
        Wing area $S$ [m$^2$] & 68.61 & 55.49 \\
        Wing span $b$ [m] & 27.05 & 27.05 \\
        Mean aerodynamic chord $\bar{c}$ [m] & 2.617 & 2.117 \\
        Aspect ratio & 10.66 & 13.19 \\
        Wing loading [kg/m$^2$] & 360 & 440 \\
        Vertical tail volume coefficient $c_{VT}$ & 0.1249 & 0.1186 \\
        Vertical tail area $S_{VT}$ [m$^2$] & 20.10 & 13.62 \\
        Vertical tail lever arm $l_{VT}$ [m] & 11.53 & 13.07 \\
        Horizontal tail volume coefficient & 1.065 & 1.065 \\
        \midrule
        \multicolumn{3}{l}{\textit{Mass and inertia}} \\
        Take-off mass [kg] & 24\,700 & 24\,416 \\
        Operating empty mass [kg] & 17\,478 & 17\,282 \\
        Fuel mass (LH$_2$) [kg] & 501 & 414 \\
        $J_{xx}$ [$10^3$ kg\,m$^2$] & 136.8 & 126.0 \\
        $J_{yy}$ [$10^3$ kg\,m$^2$] & 648.2 & 430.2 \\
        $J_{zz}$ [$10^3$ kg\,m$^2$] & 763.4 & 540.6 \\
        $J_{xz}$ [$10^3$ kg\,m$^2$] & $-17.1$ & $-11.5$ \\
        \midrule
        \multicolumn{3}{l}{\textit{Propulsion}} \\
        Number of propellers $n_T$ & 2 & 8 \\
        Rated power per unit [kW] & 2303 & 569 \\
        Propeller diameter [m] & 4.15 & 3.16 \\
        Spanwise positions $|y_i|$ [m] & 5.40 & 3.76, 7.01, \\
        & & 10.27, 13.53 \\
        \midrule
        \multicolumn{3}{l}{\textit{Trim at $M = 0.20$, sea level}} \\
        Angle of attack $\alpha_0$ [deg] & 5.86 & 8.28 \\
        Lift coefficient $C_L$ & 1.244 & 1.521 \\
        Elevator deflection [deg] & $-10.70$ & $-14.02$ \\
        Thrust per propeller $T_0$ [kN] & 7.73 & 1.73 \\
        \bottomrule
    \end{tabular}
\end{table}

Both aircraft carry the same control surfaces: ailerons with 25\,\% chord, an elevator and a rudder with 35\,\% chord, and trailing-edge flaps. The deflection limits are $\pm 14^\circ$ for the ailerons, $-25^\circ$ to $+20^\circ$ for the elevator, $\pm 25^\circ$ for the rudder, and $0^\circ$ to $40^\circ$ for the flaps. The moment sets are evaluated at $M = 0.20$ at sea level, that is, $V = 68.1$~m/s and $\bar{q} = 2837$~Pa, with the flaps at $15^\circ$.

\begin{table}
    \centering
    \caption{Control derivatives at trim ($M = 0.20$, sea level), in $10^{-3}$ per degree.}\label{tab:control_derivatives}
    \begin{tabular}{llrrrrr}
        \toprule
        & & $C_{L_\delta}$ & $C_{Y_\delta}$ & $C_{l_\delta}$ & $C_{m_\delta}$ & $C_{n_\delta}$ \\
        \midrule
        Reference & Flap     & 35.80 & 0.00    & 0.00    & $-1.72$  & 0.00 \\
                  & Aileron  & 0.00  & 0.00    & 3.46    & 0.00     & 0.18 \\
                  & Rudder   & 0.00  & $-4.50$ & $-0.53$ & 0.00     & 2.36 \\
                  & Elevator & 8.48  & 0.00    & 0.00    & $-45.12$ & 0.00 \\
        \midrule
        DEP       & Flap     & 37.42 & 0.00    & 0.00    & $-4.13$  & 0.00 \\
                  & Aileron  & 0.00  & 0.07    & 3.67    & 0.00     & 0.25 \\
                  & Rudder   & 0.00  & $-3.56$ & $-0.35$ & 0.00     & 2.00 \\
                  & Elevator & 6.28  & 0.00    & 0.00    & $-44.22$ & 0.00 \\
        \bottomrule
    \end{tabular}
\end{table}

\begin{table*}[t]
    \centering
    \caption{Vertical tail variation of the DEP aircraft.}\label{tab:sweep}
    \begin{tabular}{cccccccccc}
        \toprule
        $c_{VT}$ & $S_{VT}$ [m$^2$] & Take-off mass [kg] & $C_{n_\beta}$ & Roll & Pitch & Yaw & Violated directions & $\varepsilon$ \\
        \midrule
        0.069 & 8.09  & 24\,309 & 0.002 & 0.98 ($-0.3$) & 0.94 ($-0.8$) & 1.62 ($+5.2$)  & 32\,\% & $> 0.999$ \\
        0.094 & 10.88 & 24\,365 & 0.040 & 1.01 ($+0.2$) & 0.92 ($-1.1$) & 1.81 ($+6.1$)  & 24\,\% & $> 0.999$ \\
        0.119 & 13.62 & 24\,416 & 0.079 & 1.05 ($+0.7$) & 0.90 ($-1.4$) & 2.00 ($+6.7$)  & 23\,\% & $> 0.999$ \\
        0.144 & 16.32 & 24\,464 & 0.118 & 1.08 ($+1.2$) & 0.88 ($-1.7$) & 2.19 ($+7.1$)  & 22\,\% & $> 0.999$ \\
        \bottomrule
    \end{tabular}
\end{table*}

\begin{figure}[h]
    \centering
    \includegraphics[width=\columnwidth]{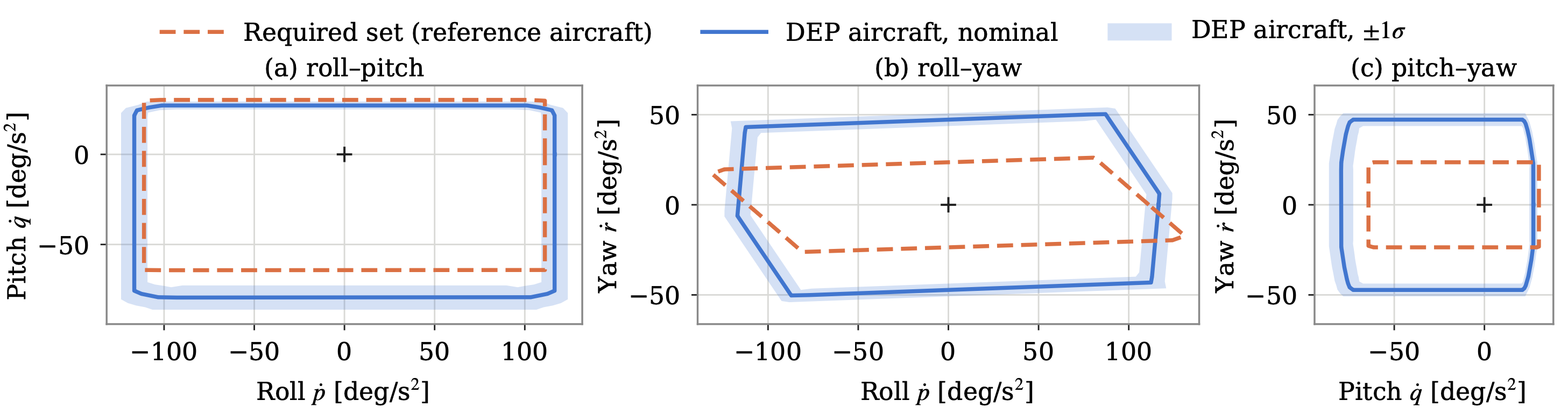}
    \caption{Cuts of the required moment set (reference aircraft) and of the attainable moment set of the DEP aircraft with its $\pm 1\sigma$ band through the coordinate planes, $c_{VT} = 0.119$, $M = 0.20$ at sea level.}\label{fig:pams_cuts}
\end{figure}

Fig.~\ref{fig:pams_cuts} shows the cuts of the required set and of the attainable set of the DEP aircraft through the three coordinate planes for $c_{VT} = 0.119$. In yaw, the DEP aircraft attains $47$~deg/s$^2$ against a requirement of $24$~deg/s$^2$, twice the authority of the reference aircraft although its vertical tail is about one third smaller, which reflects the contribution of differential thrust. In roll, the attainable radius of $117$~deg/s$^2$ exceeds the required $111$~deg/s$^2$ by less than one standard deviation. In pitch, the set is asymmetric: nose-down authority exceeds the requirement by 24\,\%, whereas nose-up authority reaches only 90\,\% of it. This asymmetry follows from the trim state in Table~\ref{tab:aircraft}, where the DEP aircraft has already used more nose-up elevator than the reference aircraft. The roll--yaw cut further shows that the required set protrudes from the attainable set at the two corners where roll is combined with yaw of opposite sign.

\subsection{Bayesian Optimization of the Tail Size}\label{subsec:bo_results}

The optimization of Section~\ref{sec:bo} is run with $c_{VT} \in [0.068, 0.300]$, $C_{n_\beta,\min} = 0.06$, and $\alpha_{\max} = 10^\circ$. With the required set of the reference aircraft at full size, none of the evaluated designs is feasible: the coverage constraint is violated by two to three standard deviations over the whole range, in directions that the vertical tail does not influence. To study how the optimizer uses the controllability signal when feasible designs exist, the required set is scaled by a factor of $0.5$ about the trim point for the results in this section.

\begin{figure}[t]
    \centering
    \includegraphics[width=\columnwidth]{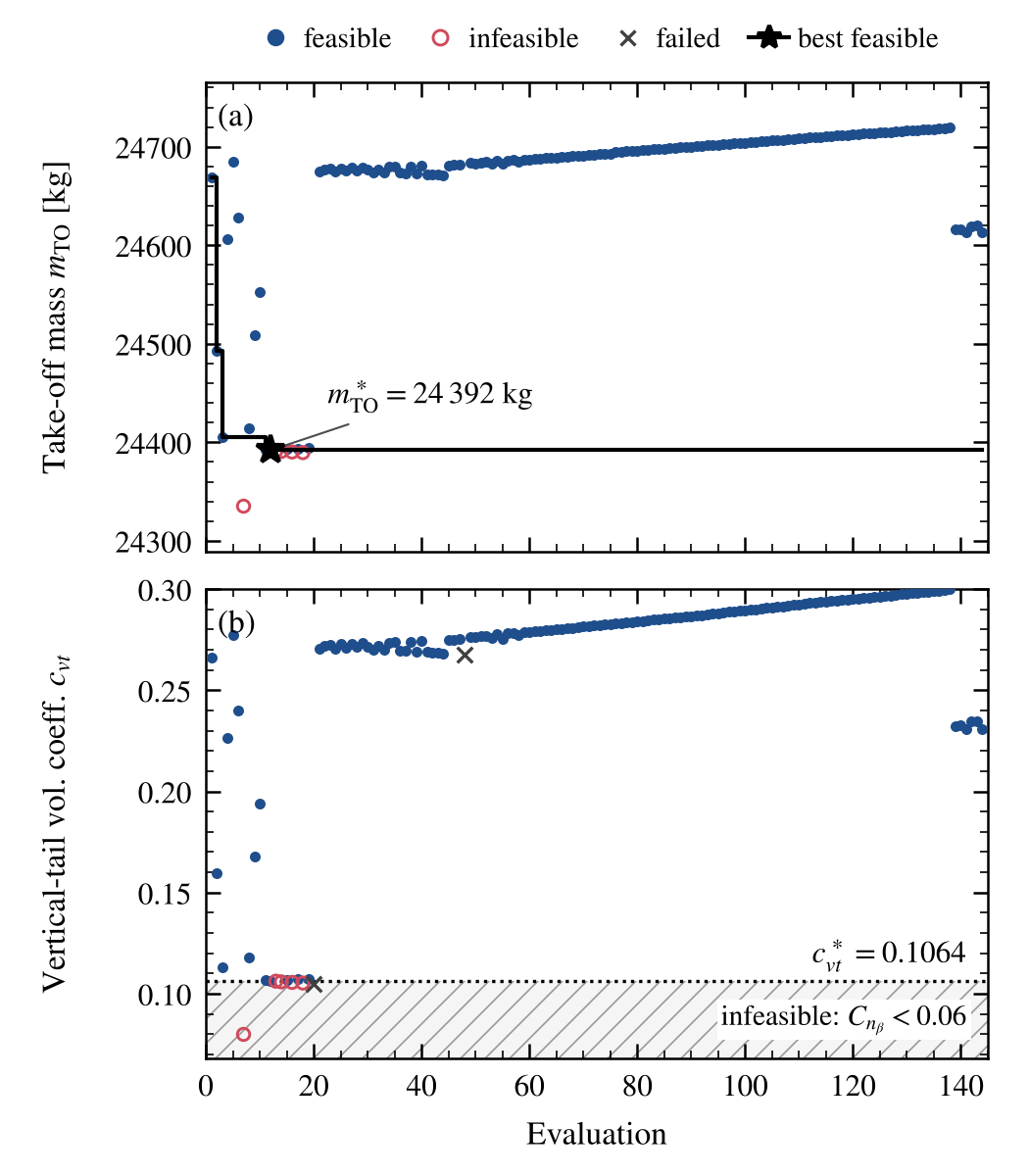}
    \caption{History of the Bayesian optimization with the required set. (a) Take-off mass of the evaluated designs and best feasible value. (b) Evaluated vertical tail volume coefficients; designs in the hatched region violate the directional stability constraint.}
    \label{fig:bo_history}
\end{figure}

Fig.~\ref{fig:bo_history} shows the history of the run, which starts from ten designs of a Latin hypercube sample. Nine of them are feasible, the lightest at $c_{VT} = 0.113$ with $24\,406$~kg. The first two guided evaluations move directly to the boundary of the feasible region and find the best design at evaluation~12,
\begin{equation}
    c_{VT}^\star = 0.1064, \qquad m_{TO}^\star = 24\,392~\mathrm{kg}.
    \label{eq:bo_result}
\end{equation}
The following evaluations are placed within $\pm 0.001$ of this value, on both sides of the boundary, and do not improve it further. At the optimum the directional stability constraint is active, $C_{n_\beta} = 0.060$, whereas the coverage constraint is satisfied with a margin of $5.6$ standard deviations and a violation probability of zero, and the trim angle of attack of $8.3^\circ$ is well below its limit. Over all evaluated designs the coverage margin stays between $4.6$ and $5.8$ standard deviations. With the scaled requirement the control authority therefore does not limit the tail, and the optimum is the smallest tail that keeps the aircraft directionally stable. Its area of $12.3$~m$^2$ is 39\,\% smaller than the $20.1$~m$^2$ of the reference aircraft. The take-off mass varies almost linearly with the tail volume coefficient, by about $1.7$~kg per $0.001$, so that the mass benefit of the smaller tail is modest.

Two sizing runs did not complete, at evaluations 20 and 48, and are recorded as failed designs. The first of them lies at $c_{VT} = 0.105$, directly next to the optimum. From this point on the search leaves the boundary and samples the upper part of the range, between $c_{VT} = 0.27$ and $0.30$, without finding a lighter feasible design.

\section{Conclusion}\label{sec:conclusion}
This paper introduced a controllability measure for integrated aircraft design. For every design of the sizing loop, the attainable moment set is built from the low-fidelity models of that loop, with the propulsion units of a distributed electric propulsion aircraft included as control effectors, and the uncertainty of these models is propagated to its boundary. The resulting probabilistic attainable moment set is compared with a required moment set that is taken from the conventional reference aircraft, so that no explicit moment requirements are needed at this stage. The comparison returns a margin for every direction of the moment space and the probability of violating the requirement, and this signal was coupled to the sizing loop through a constrained Bayesian optimization of the vertical tail. The application to a regional aircraft with eight propellers showed that the measure separates the axes on which a configuration has sufficient authority from those on which it falls short, and that it identifies whether the chosen design variable can influence the limiting axis: with a relaxed requirement the tail is sized by directional stability, whereas the full requirement is not covered in directions that the vertical tail does not affect.

This last observation defines the next step. The design variables will be extended from the tail size to the size and the location of the effectors themselves, that is, the dimensions of the control surfaces and the spanwise positions of the propulsion units, so that the optimization can act on the axes that limit the design. Further work will also consider additional flight conditions and failure cases.







\section*{ACKNOWLEDGMENT}
This work was funded by the Deutsche Forschungsgemeinschaft
(DFG, German Research Foundation) - Project ID 498601949 - TRR
364 SynTrac

\bibliographystyle{IEEEtran.bst}
\bibliography{IEEEexample.bib}

\end{document}